\documentclass[12pt]{article}
\usepackage{epsf}
\usepackage{graphicx}
\usepackage{natbib}
\usepackage[T2A,T1]{fontenc}
\usepackage{journals_joge}

\newcommand{\med}{\mathop{\mathrm{median}}\displaylimits}

\begin{document}

\title{Comparison of Different Estimates of the Accuracy of Forecasts
of the Earth's Rotation Parameters}
\author{Z. M. Malkin$^1$, V. M. Tissen$^{2,3}$ \\ $^1$Pulkovo Observatory, St. Petersburg, Russia,\\ e-mail: malkin@gaoran.ru \\
  $^2$West-Siberian branch of the FSUE All-Russian Research Institute\\ of Physical Technical and Radio Engineering Measurements,\\ Novosibirsk, Russia \\
  $^3$Siberian State University of Geosystems and Technologies,\\ Novosibirsk, Russia}
\date{\vskip -2em}

\maketitle

\begin{abstract}
Improvement of the prediction accuracy of the Earth's rotation parameters (ERP) is one of the
main problems of applied astrometry. In order to solve this problem, various approaches are used and in order
to select the best one, comparison of the accuracy of the forecasts obtained by different methods at different
analysis centers are often carried out. In such comparisons, various statistical estimates of the forecast
errors are used, based on the analysis of the differences between the predicted and final ERP values. In this
paper, we compare several prediction accuracy estimates, such as root mean square error, mean error, median error,
and maximum error. It is shown that a direct relationship between the estimates of the forecast accuracy obtained by
these methods does not always exist. Therefore, in order to obtain the most informative results of comparison
of the accuracy of different forecast methods, it is recommended to use several estimates together in the studies
dealing with comparison of the series of ERP forecasts, especially short-term ones.
\end{abstract}


\section{Introduction}

Earth's rotation parameters (ERP) are transformation
parameters between the Earth's terrestrial coordinate system,
in which the positions and velocities of the objects at the
Earth's surface and close to the latter are set and the
celestial coordinate system, in which the positions and
velocities of astronomical objects are set. High-precision
ERP values are necessary for solving many scientific
and practical problems of the astronomy, geodesy,
terrestrial and space navigation, telecommunications,
synchronization of remote time scales, satellite sensing
of the Earth's surface and atmosphere and other
applications, in particular, the ones related to the
range of problems solved by the fundamental
Position-Navigation-Timing  system \citep{Finkelstein2005}.
To ensure operations in real time, it is necessary to forecast ERP
with the highest possible accuracy. Therefore, improving
the accuracy of determining and predicting ERP is
one of the main tasks of applied astrometry. To solve it, different
authors develop various forecasting methods.
Comparison of the forecast accuracy of ERP, obtained
by different authors and methods, is a traditional task.
In particular, one can single out special campaigns
organized by the International Earth Rotation and
Reference Systems Service (IERS) \citep{Kalarus2010,Kosek2011}.

The literature describes several different statistical
estimates of the accuracy of forecasts on the basis of
which the series of forecasts obtained by different
authors and methods are compared. All of them are
based on the analysis of the differences between the
predicted and final parameters. Initially, the r.m.s.
values of these differences were usually used \citep{McCarthy1991}.
This estimate still continues to be widely used.
Later, the mean forecast error also became frequently used \citep{Kosek1998}.
In \citet{Malkin1996}, it was suggested to additionally use the maximum
forecast error, which can be considered as an
estimate of the guaranteed error, which is important
for some practical applications. In addition to these
three estimates, we also considered the median forecast
error, which is one of the robust estimates that are
not affected by random good or bad forecasts, unlike
the rest of the applied estimates. Methodology for
computation of these estimates and the results of their
application to the real forecasts of ERP are presented
below.

\section{Comparison of estimates of ERP prediction errors}

We have compared four estimates of the accuracy
of the ERP forecast using the real forecasts
of the Earth's pole coordinates $X_p$ and $Y_p$
and universal time UT1 computed in 2011--2020 at three
ERP analysis centers. The first series of forecasts is
calculated at the IERS Rapid Service/Prediction Center
operating at the U.S. Naval Observatory (USNO) \citep{IERSAR2018}.
These forecasts are published in
the daily IERS Bulletin A\footnote{https://datacenter.iers.org/availableVersions.php?id=6}
and hereinafter referred to as BA.
The second series of forecasts is calculated by
the Center for Consolidated Processing and Determination
of ERP of the Main Metrological Center of the
State Service for Time, Frequency and Determination
of the Earth's Rotation Parameters (SSTF) \cite{Kaufman2010},
which is a subdivision of the All-Russian Research
Institute of Physical, Technical and Radio Engineering
Measurements (VNIIFTRI). These forecasts are
published in the daily Q\footnote{ftp://ftp.vniiftri.ru/Out\_data/Bul\_rus\_Q/}
bulletins and are hereafter
referred to as BQ. The third series of forecasts was calculated
by Tissen, hereinafter these forecasts are
designated as VT. The methodology for calculation of
these forecasts is presented in \citet{Tissen2009,Tissen2014}.
The fourth series of forecasts was calculated by Z.M. Malkin. Hereinafter,
these forecasts are designated as ZM. To calculate
the ZM predictions, the technique described in \cite{Malkin1996}
was used, with small variations in the model parameters.

Thus, we processed four series of forecasts, which
were calculated daily in real time for a ten-year period
from January 1, 2011 to December 31, 2020. In total,
over this period, one could expect 3653 forecasts for
each series. However, for various reasons, not all forecasts
got into our database. In fact, 3626 BA forecasts,
3653 BQ forecasts, 3258 VT forecasts, and 3626 ZM
forecasts were collected. All of them, without exceptions,
were used in the calculations, the results of
which are presented below.

In the above-mentioned processing centers, the
different duration forecasts are calculated. In this
study, we used forecasts with a duration of 30 days,
which corresponds to the length of the BQ forecast,
the shortest of those considered.

When comparing the predicted ERP values with
the final ones, in order to maintain the rigor of the
comparison, one should consider which ERP series
was predicted (actually extrapolated), and the final
values of just this series should be used to calculate
the forecast errors. Predictions of BA, VT, and ZM
extrapolate the USNO ERP series finals.all\footnote{https://datacenter.iers.org/eop.php}.
Therefore, for
these series of forecasts, the comparison was made
with this series.
The BQ forecasts were compared with the final SSTF series
gs\_pvz.dat\footnote{ftp://ftp.vniiftri.ru/Out\_data/EOP\_series/gs\_pvz.dat}

The forecast errors for each analysis center (forecast series) were determined as follows.
For each of them, the differences between the predicted ERP values
and the final value of the series $d_{ij}$ were calculated,
where $i$ is the forecast number in the set of forecasts of
the given center, and $j$ is the number of days (length)
of the forecast, $j$=$1 \ldots 30$.
Next, the following statistics were calculated for each prediction series and the forecast length:

root mean square error
\begin{equation}
RMS_j = \sqrt{\frac{\sum\limits_{i=1}^n d_{ij}^2}{n}} \,,
\end{equation}

average absolute error
\begin{equation}
MAE_j = \frac{\sum\limits_{i=1}^n |d_{ij}|}{n} \,,
\end{equation}

median absolute error
\begin{equation}
MedAE_j = \med\limits_{i=1 \ldots n} \, |d_{ij}| \,,
\end{equation}

maximum absolute error
\begin{equation}
MaxAE_j = \max\limits_{i=1 \ldots n} \, |d_{ij}| \,.
\end{equation}

It should be noted that, unlike other estimates,
$MaxAE$ does not reflect some averaged accuracy of forecasts,
but is determined by the worst prediction in the series for given $j$.

\begin{figure*}
\includegraphics[clip]{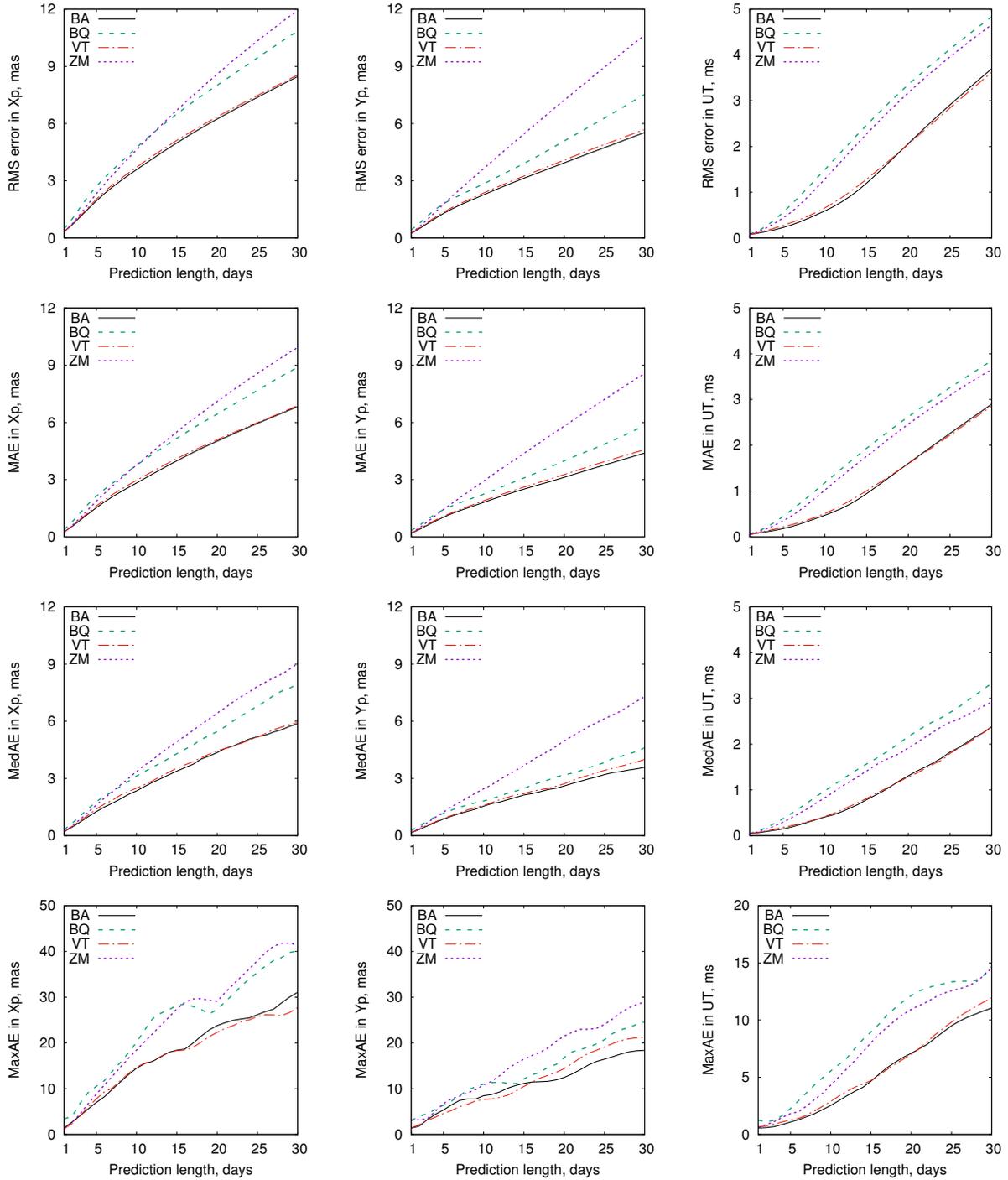}
\caption{Different estimates of the ERP forecast errors.
  The columns correspond to $X_p$, $Y_p$, and UT1, the rows correspond to $RMS$, $MAE$, $MedAE$, and $MaxAE$.}
\label{fig:errors}
\end{figure*}

The average errors calculated in this way for a
10-year period for four series of predictions of the pole
coordinates and universal time are shown in Fig.~\ref{fig:errors}.
The results obtained show that the accuracy of forecasts
obtained by different methods at different centers
of analysis, in many cases, are significantly different.

These results show that the BA and VT forecasts are the best and have approximately the same
accuracy in all cases.
A more detailed analysis reveals some advantage of one of these methods for certain
types of ERP and forecasts length intervals, but these
differences are small. The forecasts BQ and ZM are
generally much worse. For the forecasts of $X_p$ and
UT1, these two series have close accuracy. When predicting
the coordinates of the pole, the ZM results
show slightly better accuracy for short forecast length,
while BQ forecasts are significantly better for longer term
forecasts, especially for . When predicting
UT1, the ZM predictions are slightly more accurate in
all cases. It is also interesting to note that the $X_p$ forecast
accuracy was significantly worse than that for $Y_p$
for all centers.

In general, it can be noted that the comparative
assessment of the ERP forecast accuracy, which can
be made on the basis of their comparison by different
methods, is practically the same, with the exception,
in some cases, of the shortest term forecasts.

\begin{figure*}
\includegraphics[clip]{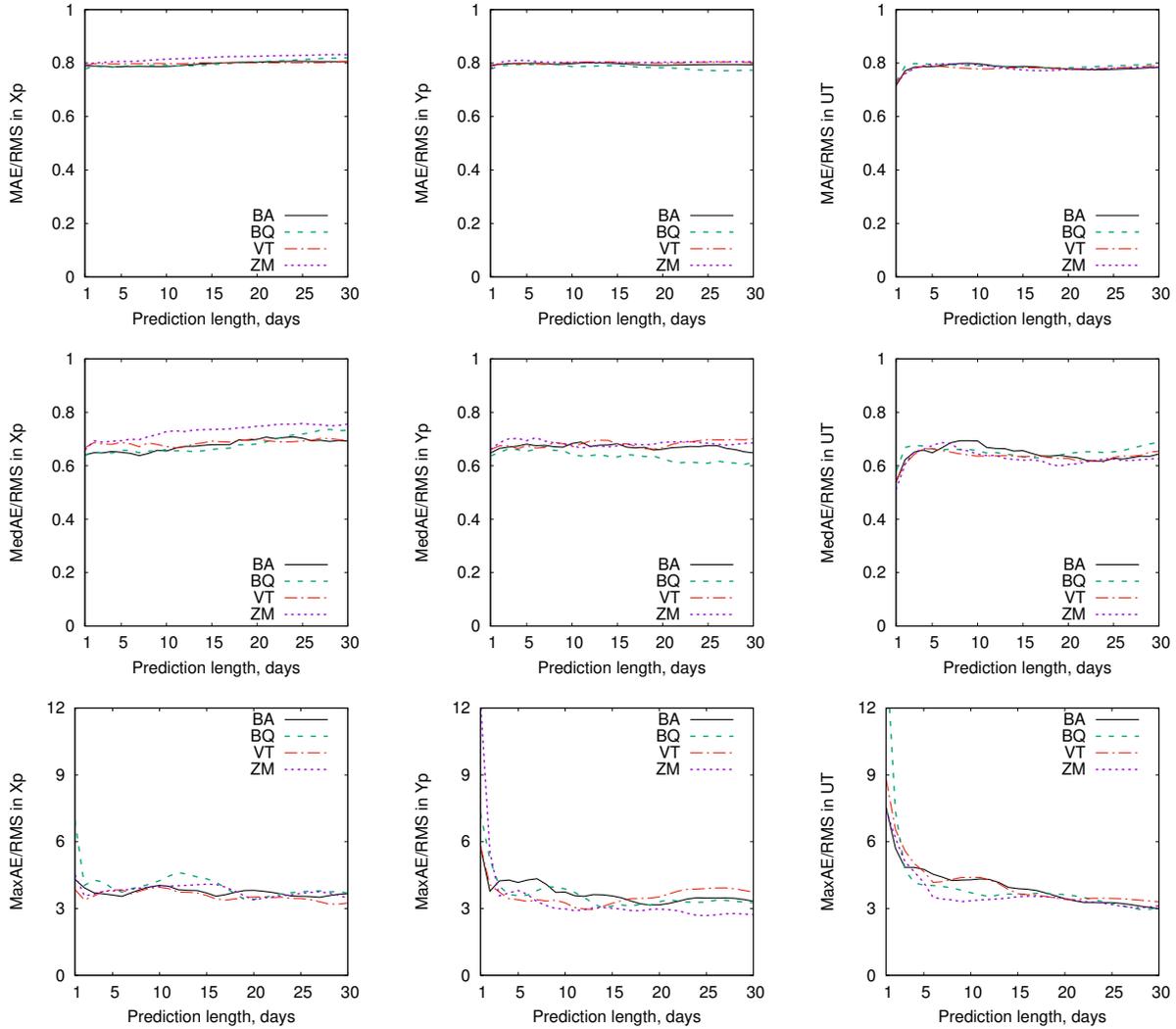}
\caption{The ratio of different estimates of ERP forecast errors.
  The columns correspond to $X_p$, $Y_p$, and UT1, the rows correspond to $MAE$/$RMS$, $MedAE$/$RMS$, and $MaxAE$/$RMS$.}
\label{fig:ratios}
\end{figure*}

Figure~\ref{fig:ratios} shows the ratios of different estimates of
the forecast errors of ERP. With the exception of the
shortest-term forecasts, the estimates obtained by different
methods for assessing the accuracy of forecasts
differ practically by a constant factor only, which is the
same for different series of forecasts and for all three
types of ERP. In all cases, the estimate $MAE$ is about
0.8 of $RMS$. This result confirms the conclusions
made in \citet{Malkin2012}, obtained for the two-year series of
USNO and VT forecasts, calculated in 2009--2011.
The $MedAE$ estimate comprises about two thirds
of $RMS$ for the forecasts longer than three to
four days. At the same time, it can be noted that the
estimate $MAE$ is much closer to $RMS$ (considering a
constant factor) than $MedAE$. The deviation of these
three estimates from a constant value is especially
noticeable for the shortest-term forecasts of universal time.

The maximum error is three to four times larger
than $RMS$ for forecasts longer than a few days for all
centers of analysis and types of ERP. At the same time,
the ratio of $MaxAe$ and $RMS$ is much higher for the
forecasts lasting one or two days. The reason for this
may be the following. As it was shown in \cite{Malkin1996}, the forecast
errors are largely affected by the ERP errors of the
last epochs in the predicted series, which have a
reduced accuracy compared to the final values calculated
after some time. Different forecasting methods
can be sensitive to this factor in various ways. Let us
recall that $MaxAE$ is determined by the most unsuccessful
forecast and this value can be considered as an
estimate of the guaranteed accuracy of the forecast in
the most unfavorable case.

\section{Conclusions}

In this paper, we have compared four different
methods for estimating the ERP forecast errors,
such as $RMS$, $MAE$, $MedAE$, and $MaxAE$.
Calculations which were made using the data of four series of real
forecasts calculated at four analysis centers
during a 10-year period from January 1, 2011 to
December 31, 2020 allow us to draw the following conclusions.

When comparing the forecasts of different centers
with a duration from several days to one month, the
$RMS$, $MAE$ and $MedAE$ estimates turned out to be
practically equivalent to within a constant factor.
Small observed differences have little effect on the
main results of comparing different series of forecasts.
Only for forecasts with very close accuracy, such as,
BA and VT, the use of one or another estimate can
show a slight advantage of one or the other series,
which is interesting theoretically, but may be of little
value for practice.

On the contrary, for short forecasts, different estimates
of the forecast accuracy can give significantly
different results. The difference in estimates is especially
large for forecasts with a duration of one or two
days, which are critical for many practical applications,
for example, related to ephemeris-time support
of global navigation satellite systems (GNSS).

The same applies to the comparative estimate of
the maximum forecast error in different centers, which
practically differ by a constant factor only, unless there
are no grossly erroneous forecasts. If the series contains
grossly erroneous forecasts, the application of the $MedAE$ estimate
allows one to obtain an estimate of the forecast
accuracy of the given series, not distorted by the
influence of individual forecasts with abnormally large
errors.

In general, it can be said that various criteria for
assessing the accuracy of forecasts of ERP are useful to
complement each other and their combined application
allows for the most complete comprehensive
comparative assessment of the accuracy of various
series (methods) of forecasting. Therefore, it seems
useful in the studies devoted to comparisons of the ERP,
to give all the above-mentioned (and, possibly, some
additional) estimates of the forecast errors in order to
provide complete information to different consumers
with their specific requirements. This is especially
important for short forecasts when various estimates of
forecast errors can differ significantly.

\bibliography{eop_pred_assess}
\bibliographystyle{joge}

\end{document}